# Droplet penetration through an inclined mesh

Long Xu[1], Shaoqiang Zong[1], Jiguang Hao[1]*, and J. M. Floryan[2]
[1]School of Aerospace Engineering, Beijing Institute of Technology, Beijing 100081, China
[2]Department of Mechanical and Materials Engineering, Western University, London, Ontario N6A 5B9, Canada

*Droplets with different Weber numbers We impacting meshes at various inclination angles α were investigated using high-speed photography. It was found that the droplet mesh penetration can be completely suppressed by inclining the mesh. The phase diagrams in the (We, α)–plane determining the expected type of penetration have been determined experimentally for meshes of various structures. It was shown that the Weber numbers for transition between no-penetration and incomplete penetration as well as for transition between incomplete penetration and complete penetration increase monotonically with α. A simple model for predicting the transition thresholds is proposed and is validated by comparisons with experimental results. It is shown that both the inclination angle and the mesh open area fraction determine the type of penetration.*

**1. Introduction**

Droplet impacts on meshes have recently attracted attention due to their importance in a range of applications including spraying[1-3], printing[4], fog collection[5-7], two-phase separation[8-11], water-proof clothing development[12], and the use of face masks to prevent virus transmission[13-15]. Following impact with high Weber numbers, the droplet penetrates through the mesh and forms liquid fingers underneath the mesh which eventually break up into secondary droplets - this process is referred to as complete penetration. The fingers below the mesh may eventually retract to the mesh surface for intermediate Weber numbers - this process is referred to as incomplete penetration. For low Weber numbers, the droplet does not penetrate through the mesh - this process is referred to as no-penetration[1-3].

Droplet penetration through a mesh may produce desired as well as undesired secondary droplets. For applications like water-proof clothing, and the development and use of masks to prevent virus transmission, the secondary droplets reduce the efficacy of the wears[16], while for spraying and printing, secondary droplets are desired. In order to control the generation of secondary droplets, it is important to determine the transition Weber number $We_{p1}$ from no-penetration to incomplete penetration as well as the transition Weber number $We_{p2}$ from incomplete penetration to complete penetration[17]. For a specific droplet, two threshold speeds of $U_{p1}$ and $U_{p2}$ can also be used to distinguish such transitions[18,19].

The analysis of droplets impacting meshes orthogonally attracted much attention during the last two decades owing to the development of high-speed photography[20]. Lorenceau and Quéré[18] studied droplet impact on a single hole, defined $U_{p2}$, and proposed a model to predict penetration based on the balance between the dynamic pressure and the capillary pressure. Similar modelling was used to study the effects of mesh materials[21-25], mesh wettability[19,26-32], mesh morphology[33-35], mesh prewetting[13,17], and droplet properties[36-40] on $U_{p1}$ and $U_{p2}$. Models[17,19,27,41] used to predict either the threshold speeds or Weber numbers were proposed for a variety of conditions. Although the goal of the studies was to improve the efficiency of creation of secondary droplets for applications like spraying, detailed studies of the secondary droplets started quite recently[1,3,13,14,28,42], focusing on the diameters, the ejection angle, and the transmitted mass both through a single hole as well as through multiple holes forming a mesh.

In applications, few impacts are orthogonal. Although droplet impacts on inclined solid surfaces attracted much attention recently[43-49], surprisingly, no attention was paid to inclined impacts on meshes. Here, we experimentally demonstrate that both incomplete penetration as well as complete penetration can be entirely suppressed by inclining the impacted mesh. The transition $We_{p1}$ from no-penetration to incomplete penetration and $We_{p2}$ from incomplete penetration to complete penetration increase monotonically with an increase of the inclination angle $α$. Here, the Weber number is defined as $We = \rho U_0^2 R/\sigma$, with $\rho$, $U_0$, $R$, $\sigma$ standing for the liquid density, the droplet impact velocity, the droplet radius, and the liquid surface tension, respectively. A modified theoretical model is proposed to describe the mechanisms responsible for these effects and is validated through comparison with the experimental data. It is shown that both the mesh morphology as well as the inclination angle $\alpha$ determine the conditions leading to a change of the type of penetration.

## 2. Experimental set up

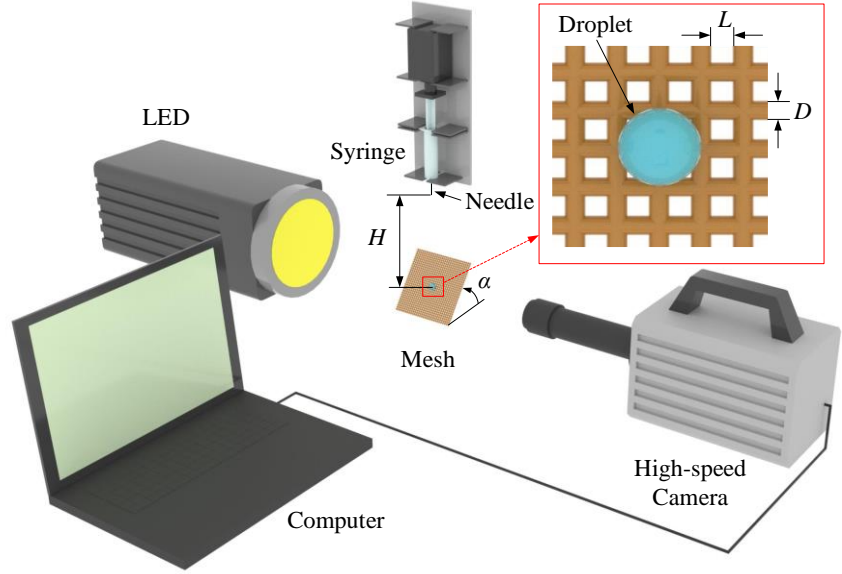

**Fig. 1.** Sketch of the experimental setup. $\alpha$ is the inclination angle. The insert illustrates the mesh structure with $L$ and $D$ denoting the pore size and the wire diameter, respectively.

As shown in Fig. 1, a flat-tipped needle driven by a syringe pump was used to create droplets of deionized water with $\rho$ = 998.7kg m$^{-3}$, $\sigma$ = 72.9mN m$^{-1}$ and viscosity coefficient $\mu$ = 1.0 mPa s[17]. The droplet radius $R$ = 1.35 ± 0.05mm was close to the capillary length of water, thus its shape oscillations preceding the impact were negligible. By adjusting the release height $H$, $U_0$ was varied from 0.44 m/s to 3.13 m/s, with $We$ varying from 3.6 to 181.2, and $Re = \rho U_0 R/\mu$ varying from 593 to 4220; this velocity was measured experimentally using image processing.

Three commercial meshes with mesh numbers $N_m$ of 60, 100, and 120 were used with the pore size $L$ and the wire diameter $D$ shown in Table 1. The open area fraction $\phi = [L/(L+D)]^2$ refers to the ratio of the pore area of the mesh to the area of the whole mesh. These meshes were placed on a rotary table whose inclination angle $\alpha$ was varied in the range 0 - 90° with a precision of ± 0.1° [45].

A Photron Nova S12 high-speed camera was used to record the impact process at a rate of 20,000 fps and with a spatial resolution of 18.75 μm/pixel. LEDs combined with a diffuser were used to provide illumination.

**Table 1** The geometric information of the tested meshes.

| Mesh number $N_m$ | The pore size $L$ (μm) | The wire diameter $D$ (μm) | The open area fraction $\phi$ |
|---|---|---|---|
| 60 | 250 | 160 | 0.372 |
| 100 | 150 | 100 | 0.360 |
| 120 | 125 | 90 | 0.338 |

## 3. Results

Figure 2 illustrates the penetration process of a droplet impacting meshes with various inclination angles at $We$ = 44.7. Each column represents a different nondimensional time $t = T/(R/V_0)$ with $T$ measured from the beginning of the impact, defined as the moment of contact between the droplet and the mesh. Each row represents an inclination angle. The mesh number $N_m$ is 100. Complete penetration occurred both for $\alpha = 0°$ (see Fig. 2(a)) and 30° (see Fig. 2(b)). The number of secondary droplets produced by the impact on a mesh with $\alpha = 0°$ was much larger than that on a mesh of $\alpha = 30°$. Figure 2(c) illustrates incomplete penetration for the impact on a mesh with $\alpha = 50°$. No-penetration occurred for the impact on a mesh with $\alpha = 70°$, see Fig. 2(d). These results clearly demonstrate that complete penetration can be suppressed into no-penetration and then into incomplete penetration by inclining the mesh.

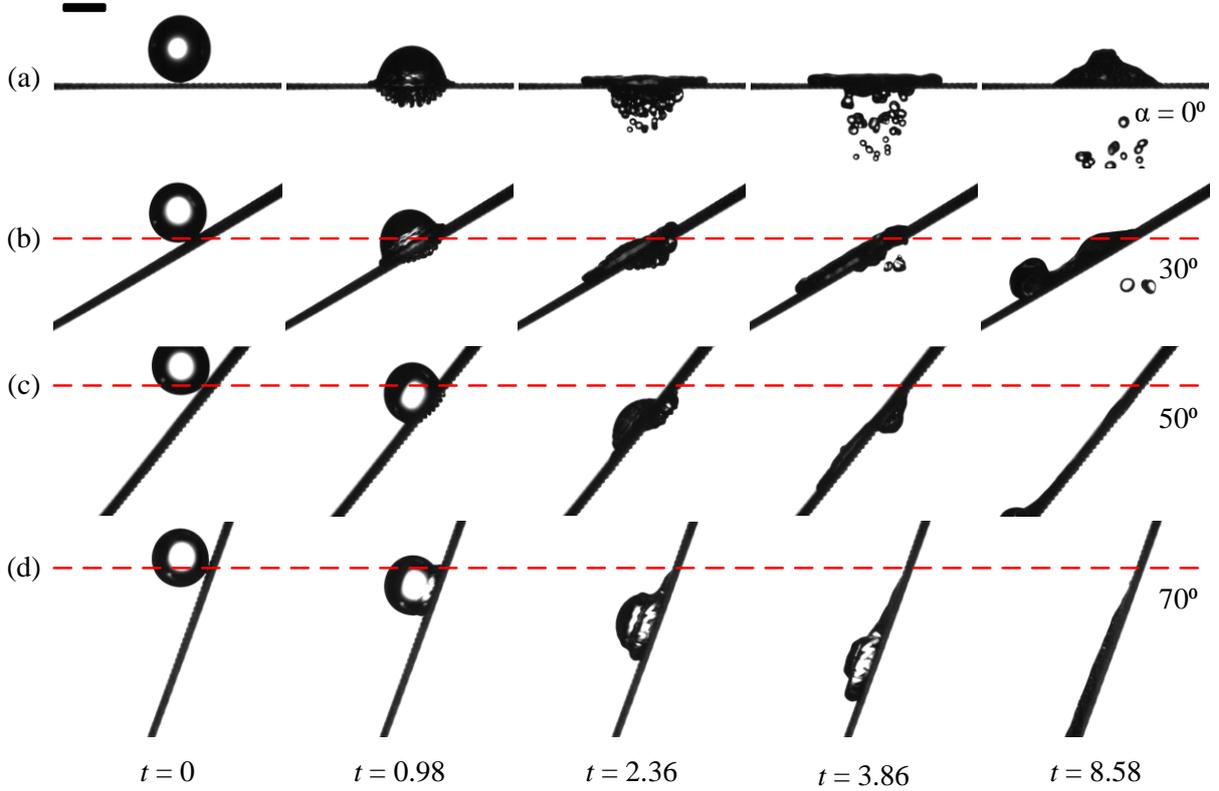

**Fig. 2.** Time evolution of a droplet impacting a mesh with $N_m =100$ with different inclination angles for $We = 44.7$. Each row represents an inclination angle. Each column represents a different time $t$. The scale bar is 2.0 mm. The red dashed lines indicate the impact point. (a) $\alpha = 0°$, complete penetration; (b) $\alpha = 30°$, complete penetration; (c) $\alpha = 50°$, incomplete penetration; (d) $\alpha = 70°$, no-penetration.

A series of experiments on droplet impacts on inclined meshes with $N_m =100$ provided the basis for the construction of the phase diagram shown in Fig. 3. Experiments were repeated three times for each set of conditions. It is shown that incomplete penetration can be suppressed into no-penetration by inclining the mesh (see the impacts with $We \approx 20$ in Fig. 3). Impacts with higher $We$'s ($We \approx 40$) in Fig.3 provide further support for the observation that complete penetration can be completely suppressed as shown in Fig. 2. Figure 3 clearly demonstrates that the threshold Weber numbers both for incomplete penetration ($We_{p1}$) and for complete penetration ($We_{p2}$) increase with an increase of the inclination angle $\alpha$.

For meshes of $N_m = 60$ and 120, the phase diagrams of impact outcomes shown in Figs. 4 and 5 show similar trends but with different threshold Weber numbers for the same inclination angle. For orthogonal impacts on meshes of various $N_m$'s (see Figs. 3, 4, and 5), both $We_{p1}$ and $We_{p2}$ increase with an increase of $N_m$, which is consistent with previous studies[30,42]. For oblique impacts, the threshold Weber numbers show the same trend.

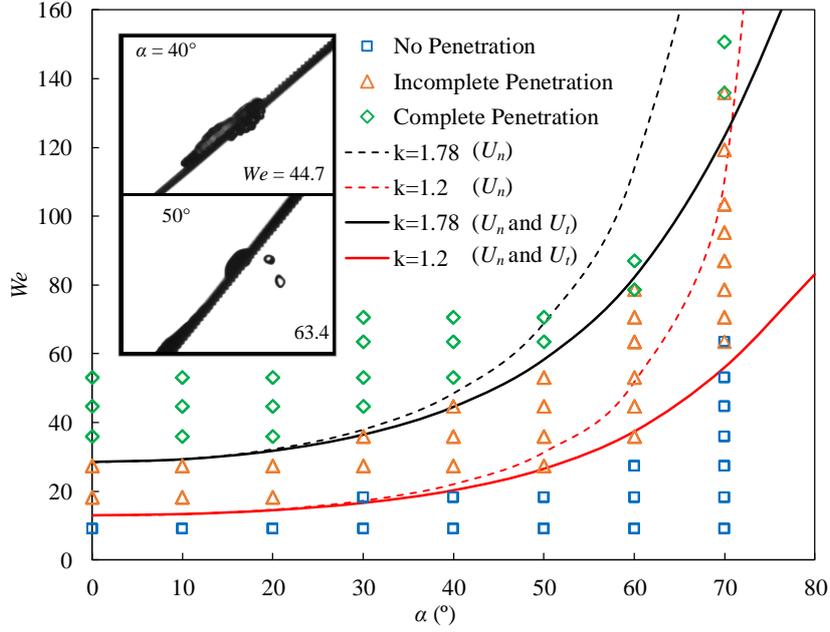

**Fig. 3.** Phase diagram illustrating the outcome of droplets' impacts on meshes with $N_m$ =100 and with varying $We$ and $\alpha$. Blue squares—no penetration; orange triangles—incomplete penetration; green diamonds—complete penetration. The red and black dashed lines represent the theoretically-predicted $We_{p1}$ and $We_{p2}$ using a simple model accounting only for the influence of the normal velocity component $U_n$. The red and black solid lines represent the theoretically-predicted $We_{p1}$ and $We_{p2}$ using model accounting for both $U_n$ and $U_t$. The inset illustrates the outcome of a droplet with the same $U_n$ = 1.19 m/s but different $We$'s impacting a mesh with $N_m$ =100 for two different inclination angles. Top: $\alpha$ = 40º, $We$ = 44.7, $t$ = 2.71, incomplete penetration. Bottom: $\alpha$ = 50º, $We$ = 63.4, $t$ = 7.41, complete penetration.

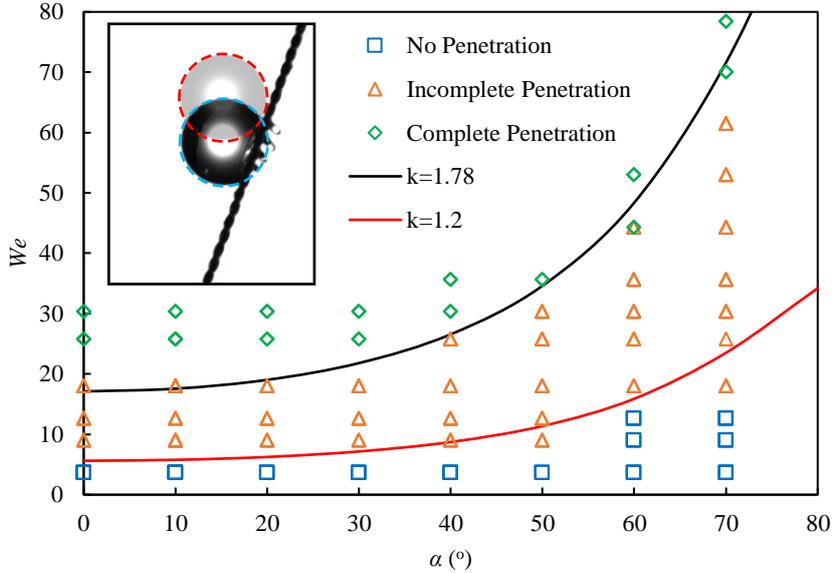

**Fig. 4.** Phase diagram illustrating the outcome of droplets' impacts on meshes with $N_m$ = 60 and with varying $We$ and $\alpha$. The red and black solid lines represent the theoretically-predicted $We_{p1}$ and $We_{p2}$ expressed by Eq. (10). The inset illustrates two steps of droplet impact for $We$ = 103, $N_m$ = 60 and $\alpha$ = 70º. The red and blue dashed lines indicate the droplet shapes and positions at $t$ = 0.0 and 1.05, respectively.

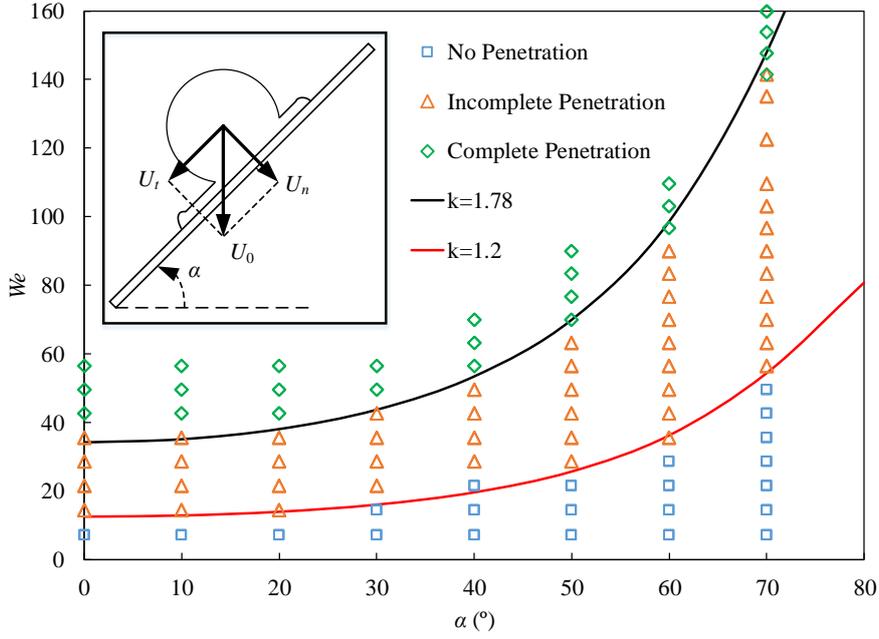

**Fig. 5.** Phase diagram illustrating the outcome of droplets' impacts on meshes with $N_m$ = 120 with varying $We$ and $\alpha$. The red and black solid lines represent the theoretically-predicted $We_{p1}$ and $We_{p2}$ expressed using Eq. (10). The inset demonstrates the decomposition of $U_0$ into the normal- ($U_n$) and tangential- ($U_t$) to-the-mesh impact velocity components, respectively.

## 4. Discussion

Previous studies[18,19,26,27] of droplets vertically impacting a mesh show that during penetration, the inertial force tends to promote penetration through the pores, while the viscous force related to flow through the pores and the capillary force resisting the formation of liquid fingers beneath the mesh prevent penetration. It has been shown that the viscous force plays a small role in preventing penetration at Reynolds numbers $Re > 100$ [18]. Since the minimum Reynolds number was $Re = 593 \gg 100$ (minimum $U_0$ of 0.44m/s) for the conditions used in the current experiments, the effects of viscous forces are neglected, with the droplet penetration through a mesh determined by the balance between the dynamic and capillary pressures.

A droplet with a mass $m \sim \rho R^3$ impacts the mesh at a speed of $U_0$ which decreases to zero after a time $\tau \sim R/U_0$. The impacted area is $S \sim R^2$. The dynamic pressure exerted by the inertia can be expressed as $P_d = mU_0/\tau S \sim \rho U_0^2$. The capillary pressure is $P_c \sim \sigma \Gamma / A$ with the open pore area being $A \approx L^2$ and the perimeter of a single pore being $\Gamma \approx 4L$. The droplet can penetrate the mesh only if $P_d > P_c$ [18,19,26,27]. The balance between $P_d$ and $P_c$ leads to

$$C \rho U_0^2 = \sigma \Gamma / A \tag{1}$$

where $C$ is a constant to be determined experimentally. This constant provides the means for predicting the threshold velocity $U_p$ required to achieve each type of penetration. The velocity $U_p$ required to achieve penetration can be evaluated as

$$U_p = \sqrt{\frac{\sigma \Gamma}{C \rho A}} = k \sqrt{\frac{\sigma}{\rho L}} \tag{2}$$

where $k = \sqrt{4/C}$ depends on the penetration type and the associated threshold velocity, i.e., either $U_{p1}$ or $U_{p2}$. Its values were determined experimentally for orthogonal impacts, i.e., $k$ = 1.2 for $U_{p1}$[18], and $k$ = 1.78 for $U_{p2}$[19].

To explain the effect of the inclination angle $\alpha$, we decompose the impact velocity $U_0$ into the normal-to-the-mesh ($U_n$) and parallel-to-the-mesh ($U_t$) components (see the inset in Fig. 5), i.e.,

$$U_n = U_0 \cos\alpha, \ U_t = U_0 \sin\alpha. \tag{3}$$

It may appear that only $U_n$ determines the penetration type. Substituting $U_n$ of (3) into (2) leads to

$$U_p = \frac{k}{\cos\alpha}\sqrt{\frac{\sigma}{\rho L}} \tag{4}$$

This relation reduces for $\alpha = 0°$ to (2) for orthogonal impacts. Substituting (4) into the Weber number, the threshold Weber number $We_p$ for a droplet impacting an inclined mesh is given as

$$We_p = \frac{\rho U_p^2 R}{\sigma} = \frac{k^2}{l \cos^2\alpha} \tag{5}$$

where $l = L/R$ is the dimensionless width of a mesh pore (see Fig.1).

The threshold Weber numbers $We_{p1}$ and $We_{p2}$ determined using (5) for $N_m = 100$ are shown in Fig. 3 using red and black dashed lines, respectively. For $\alpha < 40°$, the theoretical results agree with the experiments reasonably well. However, the deviation between the theoretically-determined and experimentally-determined threshold Weber numbers increases with a further increase of $\alpha$, indicating that the use of $U_n$ is insufficient to predict the penetration - the inset in Fig. 3 shows that impacts with the same $U_n$ produce different penetration outcomes.

Equation (3) indicates that $U_t$ increases with an increase of $\alpha$, and $U_t > U_n$ when $\alpha > 45°$. Interestingly, 45° is close to the inclination angle of 40° where the theoretical prediction based on $U_n$ begins to deviate from the experiments (see dashed lines in Fig.3). It is likely that $U_t$ plays a role in the determination of the penetration type. In addition, fingers below the mesh are nonorthogonal to the mesh for nonorthogonal impact, providing further evidence that $U_t$ must play a role in the penetration (the inset in Fig.4).

The no-slip boundary condition reduces $U_t$ at the wires to zero, while $U_t$ of the liquid in the pores remains unchanged. For simplicity, we assume that the average tangential velocity of the liquid in contact with the mesh decreases to $\phi U_t$ due to the resistance provided by the wires, where $\phi$ is the open area fraction. The normal velocity $U_n$ remains unchanged. The velocity $U_L$ of the liquid in contact with the mesh is given as

$$U_L = \sqrt{(\phi U_0 \sin\alpha)^2 + (U_0 \cos\alpha)^2} = U_0\sqrt{(\phi\sin\alpha)^2 + (\cos\alpha)^2}. \tag{6}$$

When the liquid impacts the mesh with velocity $U_L$, it produces a dynamic pressure

$$P_d = \rho U_L^2 = \rho U_0^2 \left(\phi^2 \sin^2\alpha + \cos^2\alpha\right). \tag{7}$$

Substituting (7) into (1) leads to

$$C\rho U_0^2 \left(\phi^2 \sin^2\alpha + \cos^2\alpha\right) = \sigma\Gamma/A. \tag{8}$$

The velocity $U_p$ required to achieve penetration of inclined meshes can be determined as

$$U_p = k\sqrt{\frac{\sigma}{\rho L\left(\phi^2 \sin^2\alpha + \cos^2\alpha\right)}}. \tag{9}$$

This relation reduces for orthogonal impacts ($\alpha = 0°$) to (2). Substituting (9) into the definition of the Weber number leads to the threshold Weber number $We_p$ for impacts on inclined meshes in the form of

$$We_p = \frac{\rho U_p^2 R}{\sigma} = \frac{k^2}{l\left(\phi^2 \sin^2\alpha + \cos^2\alpha\right)}. \tag{10}$$

For $N_m$ =100, Fig.3 shows $We_{p1}$ and $We_{p2}$ determined using (10) with red and black solid lines, respectively. For $\alpha < 40°$, the theoretical results using (10) are close to those using (5) and agree well with both the experiments and the previous studies of orthogonal impacts[17,19,21,27]. When $\alpha > 40°$, the theoretical results using (10) agree well with the experiments while those based on (5) do not agree even qualitatively, indicating that $U_t$ does play a role in the penetration process. For $N_m$ = 60 and 120, the theoretically-determined $We_{p1}$ and $We_{p2}$ using (10) are shown in Figs. 4 and 5 using red and black solid lines, respectively. The good agreement between the theoretical and experimental results further supports the validity of the proposed model. Equation (10) further shows that the knowledge of both the inclination angle and the mesh open area fraction are required for predicting the type of penetration.

It is shown that theoretical models not accounting for the effect of $U_t$ significantly overpredict the threshold Weber numbers at higher inclinations angles. Model accounting for both $U_n$ and $U_t$ provides a good predictive tool as verified by good agreement between experiment and predictions.

## 5. Conclusion

In conclusion, we experimentally observed that both incomplete penetration as well as complete mesh penetration by an impacting droplet can be entirely suppressed resulting in no-penetration by simply inclining the mesh. A theoretical model predicting the threshold Weber numbers $We_{p1}$ and $We_{p2}$ for droplets impacting inclined meshes is proposed and validated through comparison with the experimental data. It is shown that the inclination angle as well as the open area fraction have a decisive effect on the droplet ability to penetrate an inclined mesh.

This study was financially supported by the National Natural Science Foundation of China under Grant No. 12072032, National Key R&D Program of China under Grant No. 2018YFF0300804, and 111 Project under Grant No. B16003.

*hjgizq@bit.edu.cn